# Self-organized Growth of Graphene Nanomesh with Increased Gas Sensitivity

Matthias König[1,2,*], Günther Ruhl[2], Joerg-Martin Batke[2] and Max C. Lemme[1]

[1]University of Siegen, Department of Electrical Engineering and Computer Science, Hölderlinstr. 3, 57076 Siegen, Germany
[2]Infineon Technologies AG, 93049 Regensburg, Germany
[*]Matthias.Koenig@infineon.com

## Abstract

A bottom-up chemical vapor deposition (CVD) process for the growth of graphene nanomesh films is demonstrated. The process relies on silicon nanospheres to block nucleation sites for graphene CVD on copper substrates. These spheres are formed in a self-organized way through silicon diffusion through a 5 µm copper layer on a silicon wafer coated with 400 nm of silicon nitride. The temperature during the growth process disintegrates the $Si_3N_4$ layer and silicon atoms diffuse to the copper surface, where they form the nanospheres. After graphene nanomesh growth, the Si nanospheres can be removed by a simple hydrofluoric acid etch, leaving holes in the graphene film. The nanomesh films have been successfully transferred to different substrates, including gas sensor test structures, and verified and characterized by Auger, TEM and SEM measurements. Electrical/gas-exposure measurements show a 2-fold increase in ammonia sensitivity compared to plain graphene sensors. This improvement can be explained by a higher adsorption site density (edge sites). This new method for nanopatterned graphene is scalable, inexpensive and can be carried out in standard semiconductor industry equipment. Furthermore, the substrates are reusable.

Graphene, a carbon allotrope of the two-dimensional material class, has attracted much interest because of its extraordinary intrinsic electronic[1], mechanical[2], optoelectronic[3] and thermal[4] properties. In addition, the two-dimensional nature of the material gives rise to extreme sensitivity to its environment, suggesting applications in gas and environmental sensing[5–8]. Its remarkable properties can be tuned even further by modifying it at the nanoscale, e.g. nanoribbons that exhibit a band gap[9,10], improved contact resistance through local contact patterning[11] or enhanced light absorption in nanostructures[12]. There are several state-of-the-art methods for graphene patterning. A straight forward technique is conventional electron beam lithography[9,10], but there are also non-conventional methods based on lithographic processes like block-copolymer lithography[13–15], nanosphere lithography[16–19] and nanowire lithography[20]. Bottom-up techniques have been demonstrated to grow precise molecular-scale nanoribbons [21]. Some of these methods cannot be scaled up for industrial use, and most of them suffer from contamination issues, like residues from polymeric and inorganic species from the process chemicals, which lead to degraded carrier mobilities and random Dirac-Point shifts[22]. Independent of the technique, the resultant graphene nanostructures yield highly reactive graphene edges after patterning, with a number of possible chemical terminations. These edges may also be extremely defective or well-defined with a clear crystallographic orientation[23]. Depending on the type of application, a high amount of edge defects can be detrimental (e.g. for generating a controlled bandgap[24]) or beneficial (e.g. for electrical contacts[11]). In particular, it has been shown that gas sensitivity can be enhanced considerably in defective graphene[25] or patterned graphene meshes [13,19]. Here, we present a new bottom-up method to synthesize patterned graphene in a simple, reproducible way. We further demonstrate superior gas sensing properties of devices made with these self-organized graphene nanomeshes.

To fabricate the graphene nanomeshes, a 5 µm thick copper (Cu) film was sputter-deposited on a silicon (Si) wafer coated with 400 nm silicon nitride ($Si_3N_4$) (Figure 1a). These samples were then placed in a CVD hot wall reactor and processed at 1000°C for 10 min under hydrogen ($H_2$) atmosphere, followed by graphene growth for 10 min in $C_2H_4$ atmosphere. During this process, $Si_3N_4$ starts to decompose and Si diffuses towards the Cu surface where it forms spherical aggregates in the nanometer scale. This is shown schematically in Figure 1b). It is important to note that the Si nanosphere growth takes place already in the annealing phase, prior to the start of the graphene film growth. The areas that become occupied by Si hence locally block the subsequent catalytic graphene growth. This leads to discontinuous graphene growth only between the Si nanospheres. A similar approach was reported by Yi et al.[26], who generated the blocking sites through self-assembled colloidal silica spheres. However, it is not clear what kind of contamination is introduced into the graphene films during the reported synthesis by the Stöber method[27] and the Langmuir-Blodgett assembly. The method proposed here, in contrast, relies on standard semiconductor technology. This includes using copper-coated silicon wafers, as copper foil is quite unusual in semiconductor manufacturing, and reducing contamination issues (expected from the state of the art transfer) to a minimum. After 10 minutes of graphene growth time the samples were cooled down with a rate of 15 K/min to room temperature in hydrogen atmosphere. The Si-clusters were removed with hydrofluoric acid (HF), resulting in the graphene nanomesh structure shown schematically in Figure 1c). After HF etching, the graphene nanomesh was coated with Polymethylmethacrylate (PMMA) and the Cu was underetched with 1 mol $FeCl_3$ solution. The floating PMMA/graphene film was rinsed and picked up with a $SiO_2$-coated Si wafer. At this stage, the $Si/Si_3N_4$ substrates are reusable, standard substrate cleaning procedures and new sputter deposition of Cu will re-establish the initial conditions. The sample was heated in a UHV furnace at 400°C for 10 min to remove residual PMMA and HF. Scanning electron microscope images of transferred graphene meshes after nanosphere growth and after nanosphere removal are shown in Figure 1d) and Figure 1e), respectively. Some resultant copper surfaces with differently sized silicon nanospheres are shown in the SEM images in Figure 2a). The process conditions can clearly be tuned by growth temperature and time to adjust the nanosphere size and densities to the desired values (10-100nm). The magnification of each image was optimized to visualize the nanospheres in each process condition. The as grown samples where investigated by SEM, TEM and Auger electron spectroscopy (Figure 2b,c). Auger electron spectra (figure 2c) revealed a silicon surface concentration of 47 atomic%, corresponding to a Si/carbon surface concentration ratio of approximately 1. Details of the extraction procedure of the surface concentration are described in the methods section. The element mapping (EFTEM) of a TEM cross section in figure 2b reveals that the Si nanospheres oxidize at their surface, which corresponds with the high oxygen amount seen in the Auger electron spectrum. This enables their wet chemical removal with HF. The graphene film between the Si nanospheres is clearly visibly (marked yellow), and is not present on the Si spheres.

The proposed mechanism for the nanosphere formation was verified experimentally by measuring the diffusion constant of silicon in copper. For this purpose the same substrates as for graphene nanomesh CVD were used (Si wafers coated with 400 nm of $Si_3N_4$ and 5 µm of Cu). A temperature treatment similar to the graphene growth experiment was performed in an RTP reactor under forming gas (4% $H_2$ in $N_2$). The samples were heated with a ramp of 25 K/s, annealed at 850°C for 1, 3, 5 and 7 min intervals, and then cooled down rapidly with a rate of 25 K/s. The diffusion constant was calculated from the diffusion pair model

$$c(x,t) = 2c_0 \cdot erfc\left(\frac{x}{2\sqrt{Dt}}\right) \quad with \quad erfc(z) = 1 - \frac{2}{\sqrt{\pi}}\int_0^z \exp(-\varsigma^2)\, d\varsigma$$

where c(x,t) is the silicon concentration at the point x at time t, D is the Diffusion constant and $c_0$ is the concentration at the interface. The missing parameters c(5µm,t), i.e. the Si concentrations at the Cu surface resulting from varying anneal times, were measured by Auger electron spectroscopy (see also Methods section). The corrected Si surface concentrations vs. anneal times are summarized in Table **1**. The diffusion constant in the samples was determined to be D = $2.4 \cdot 10^{-14}$ m²/s. This is in good agreement with the reported literature value of $5 \cdot 10^{-14}$ m²/s[28]. Reducing the growth temperature even further to 800 °C slowed down the Si diffusion significantly. In fact, the Si nanosphere coverage was hardly detectable by SEM and Auger electron spectroscopy. Howsare et al. investigated barriers for graphene growth on copper in a similar configuration.[29] Their work suggests that the growth of Si nanospheres is also possible with other barriers like $SiO_2$, but with different growth conditions, attuned to the barrier materials' chemical stability in contact with Cu. Pure Si, without barriers, will most likely not work because it forms a Si/Cu alloy and Cu silicides, which prevent graphene growth.

Table 1: Si concentration vs. anneal time after attenuation correction.

| Annealing time | 1 min | 3 min | 5 min | 7 min |
|---|---|---|---|---|
| Si-surface concentration | 0% | 0 % | 16 % | 28 % |

The grown graphene nanomeshes can be transferred to arbitrary substrates after the silicon nanosphere removal by established transfer methods (as in Figure 1 d,e). In this case a common wet transfer method with a PMMA film as a support layer and $FeCl_3$ as the Cu wet etchant was used[30–33]. Fluorine residues from the HF treatment can still be detected after the transfer, but a 10 min anneal at 300°C in an UHV furnace reduces the fluorine residues below the detection limit of Auger electron spectroscopy.

Defects and edges of graphene sheets are preferred adsorption sites for gas molecules. An important issue for manufacturing graphene devices is the sensitivity towards contamination, thus we investigated the effect of amines, which are typical gaseous contamination species in semiconductor manufacturing lines, e.g. from photoresist developers. In this study we used ammonia as model test gas: 20 samples were prepared: 10 samples with graphene nanomeshes grown according to the schematic process flow in figure 1a-c and 10 samples with homogeneous graphene films, produced at a lower temperature (800°C) and growth time to avoid Si diffusion as described above. The films were transferred onto a gold meander electrode structure for electrical measurements (figure 3 a,b). The layout allows two-point and 4-point I-V measurements, but the contact resistance proved to be negligible due to the extremely long contact length. Thus, only 2-point measurements were performed. The sheet resistance of several samples (both samples) was in the range of 10 kΩ to 1 MΩ, which is expected given the high defect density. Charge carrier mobility measurements are not meaningful due to the random device geometry and unknown current paths. A back gate sweep, where the Si substrates works as the gate electrode, indicates that the devices are working like typical graphene field effect transistors (figure 3d). In a flow-through gas exposure system (figure 3c), all samples were initially exposed to 200 sccm synthetic air flow at room temperature and pressure. After 400 s, 50 ppm of ammonia was added to the synthetic air flow for 900 s, before a final pure synthetic air purge. All measurements were done at constant measurement power ($I_D \cdot V_{SD}$ = 1 mW). Some measurements hence show a low S/N ratio due to the low measurement current. Figure 3 e) compares measurements of one graphene nanomesh sensor and one graphene reference sensor. The resistance change of the devices was calculated by dividing the resistances before the start (at 400 s) and at the end (at 1200 s) of ammonia exposure:

$$R_{change} = \frac{R(t = 400s)}{R(t = 1200s)}$$

All samples showed a resistance change between 2% and 8%. Generally the nanomesh samples show an increased sensitivity towards ammonia by an average factor of 1.6 (range: 0.85…2.14). Cagliani et al. reported a more drastic difference in resistance change for lithographically etched nanomesh devices[13], but under different measurement conditions. Under comparable measurement conditions Paul et al. [19] found a sensitivity increase on lithographically etched nanomesh devices by a factor of 4.4. When analyzing the graphene egde/area ratio, which is mainly determining the gas sensitivity, by using SEM images our samples show a significantly lower ratio. Thus correcting our samples for this ratio, a sensitivity increase of factor 5.5 is found, which is in the same order of magnitude as in [19]. Additionally in this work the reference samples are grown at lower temperature which is known to yield very defective films[34]. Thus the reference samples exhibit already increased gas sensitivity.

Furthermore, an incomplete recovery of the resistance values is observed after ammonia exposure. This can be attributed to the fact that the measurements were carried out at room temperature and ambient pressure, leading to incomplete gas desorption. The resistance changes of the entire set of samples, randomly chosen from different growth runs, are summarized in figure 3f). The data was analyzed with a t-test to demonstrate the statistical significance of the difference between the two groups. The average value in the reference group is 3.54% with a standard deviation of 1.17%, while the graphene nanomesh sensors show an average of 5.66% with a standard deviation of 1.59%. The higher standard deviation in the nanomesh group can be explained by the fact that these samples have seen additional process steps with more influence sources. An F-Test ($\alpha$=0.05) shows that there is no significant difference in the standard deviations. The two-tailed P value equals 0.33%, which means that the difference between the two groups is statistically very significant using conventional criteria (i.e. a 95% confidence interval).

We demonstrated the fabrication and performance of graphene nanomesh devices through a bottom-up growth method that blocks certain growth sites on copper substrates with silicon nanospheres. These spheres are generated by diffusion of Si through a copper film at high temperatures. The diffusion mechanism was investigated by diffusion experiments using Auger electron spectroscopy measurements of the Si concentration on the Cu surface. The experimentally measured Si diffusion constant is consistent with literature. The Si nanospheres oxidize in air, which is shown in TEM cross sections, and thus can be removed by a HF wet etch. The graphene nanomesh films were transferred to large-area sensor test structures. Exposure to ammonia gas showed a factor of 1.6 increase in sensitivity compared to non-perforated reference graphene films. A commercial ammonia ZnO gas sensor is not working at room temperature. At 400 °C the ZnO has a sensitivity of 1,7 %/ppm compared to 0,16 %/ppm of the perforated sensor at room temperature [35]. The proposed bottom-up growth method is simple, scalable in size and was demonstrated with typical semiconductor manufacturing equipment. It can be used to manufacture low cost, large scale graphene nanomesh films e.g. for sensor applications. In addition, it may be utilized to improve metal-graphene contacts [11] if it can be applied locally on pre-patterned substrates.

Acknowledgements

The authors gratefully acknowledge funding from the German Federal Ministry of Education and Research (BMBF, NanoGraM, 03XP0006C). ML further acknowledges funding from the European Commission (ERC Starting Grant, InteGraDe, 307311) and the German Research Foundation (DFG, LE 2440/1-2).

Methods

*Nanomesh growth*

5 µm Cu films on 400 nm $Si_3N_4$ were used as CVD substrates. Prior to the CVD process in a laboratory reactor, the samples were annealed in hydrogen at 1000 °C to remove copper oxides. The CVD process was done at 1000 °C using an ethane/hydrogen mixture at 1 Torr. After removing the Si spheres with 3% HF solution (semiconductor grade) the samples were spin-coated with PMMA. Subsequently the Cu film was underetched with 1 mol $FeCl_3$ solution and the floating PMMA/graphene film was picked up with a $SiO_2$ wafer. Heating in a UHV furnace at 400°C for 10 min removed residual PMMA and HF. After the transfer the $Si/Si_3N_4$ substrates are reusable. Cleaning of the substrates and sputter deposition of Cu will lead to the initial conditions.

*Auger measurements*

The Si surface concentration after annealing for 1, 3, 5 and 7 min was measured with Auger electron spectroscopy. As surfaces in ambient are covered with adventitious carbon, the signal attenuation by this carbon layer has to be eliminated to calculate the correct surface composition. The extraction procedure, including elimination of the natural carbon contamination, was as follows: The signal intensities of the Auger electron peaks from carbon ($C_{KLL}$ at 275 eV), copper ($Cu_{LMM}$ at 922 eV) and silicon ($Si_{KLL}$ at 1621 eV) were multiplied with the individual sensitivity factors to calculate the surface concentrations. The $C_{KLL}$ / ($Cu_{LMM}$ + $Si_{KLL}$) intensity ratio leads to a virtual atomic concentration ratio of 50% Si on Cu which corresponds to an estimated carbon layer thickness of 1nm. The electron energy dependent attenuation in the carbon layer can be eliminated by a correction factor depending on the mean free path $\lambda_A(E_A)$ of the element specific Auger electrons in the carbon layer and the thickness of the layer $d_A$:

$$I_{real} = I_{measured}(1 - e^{-\frac{d_A}{\lambda_A(E_A)}})^{-1}$$

The inelastic electron mean free path was calculated by the NIST electron mean-free-path database v1.1 software, yielding values of 1.53 nm for the $Cu_{LMM}$ and 2.51 nm for the $Si_{KLL}$ Auger electrons. To obtain the real Cu/Si atomic ratio, the Si atomic concentration was corrected from the native $SiO_2$ layer taking the ion radii of $Si^{4+}$ (40 pm) and $O^{2-}$ (140pm) into account.

*Electric gas measurements*

The graphene films were transferred to test devices with gold meander structures (figure 3a,b). The gas exposure was controlled by mass flow controllers. The samples were exposed to gas inside a small housing, allowing fast gas exchange. I-V characteristics were measured using a Keithley 2400 source meter. In this setup, the back gate cannot be controlled during the gas measurements. However, as the measurement chamber is closed and the chip area is rather large compared to the active area, we conclude that a floating back gate should have a negligible influence.

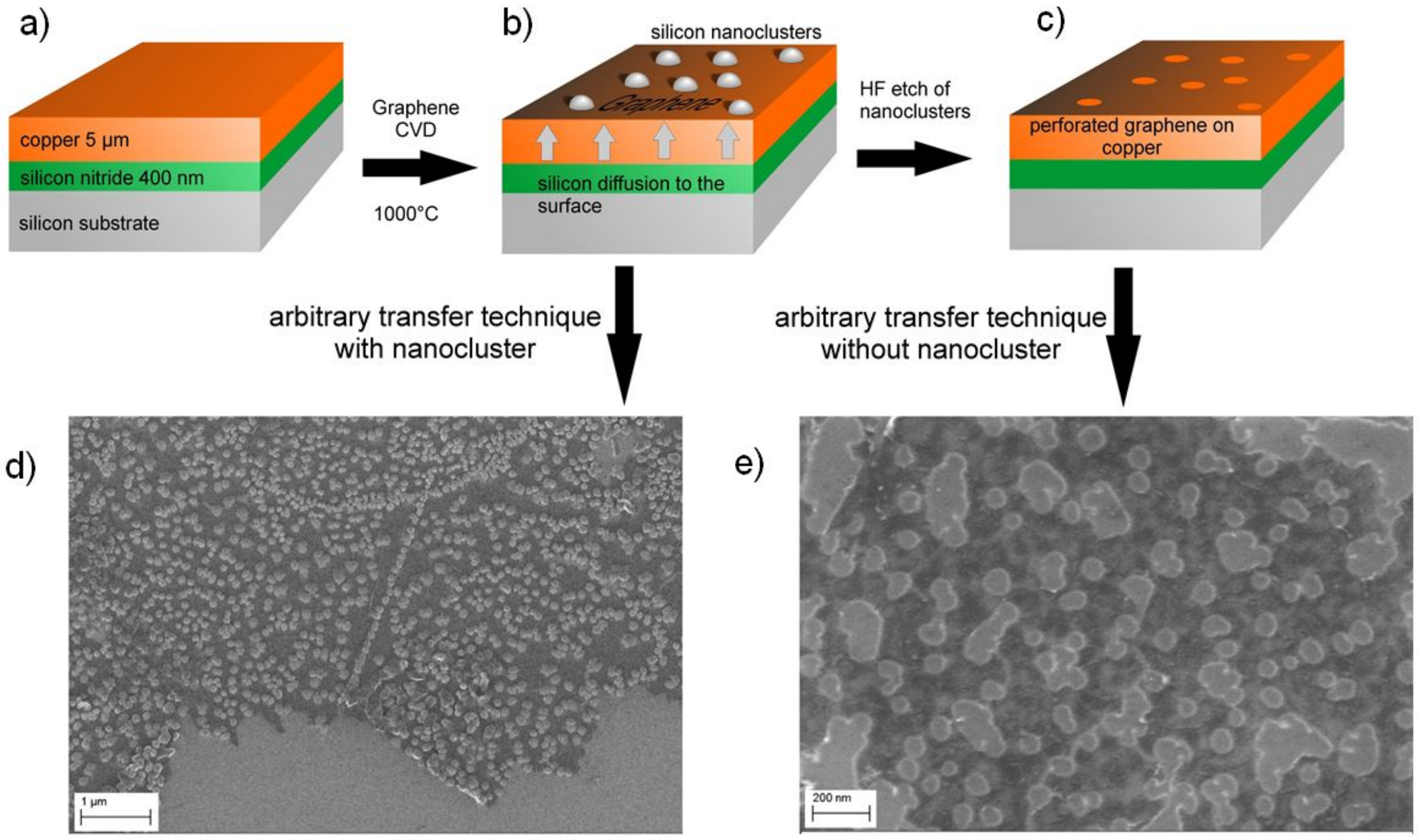

Figure 1: Fabrication scheme for bottom-up growth of graphene nanomeshes. a) Substrates consist of a 5 µm Cu film sputtered onto a silicon wafer coated with 400 nm silicon nitride. b) Pre-deposition annealing and graphene CVD is carried out at 1000 °C. During annealing, silicon diffuses from the silicon nitride/copper interface to the surface, where it forms silicon nanospheres. Graphene growth on / under the silicon nanospheres is inhibited. c) After HF etch, holes appear in the graphene layer at the former location of the silicon nanospheres. d) Scanning electron micrograph of a graphene / silicon nanosphere hybrid material transferred onto an oxidized silicon wafer. e) Scanning electron micrograph of a graphene nanomesh after HF silicon removal transferred onto nanosphere an oxidized silicon wafer.

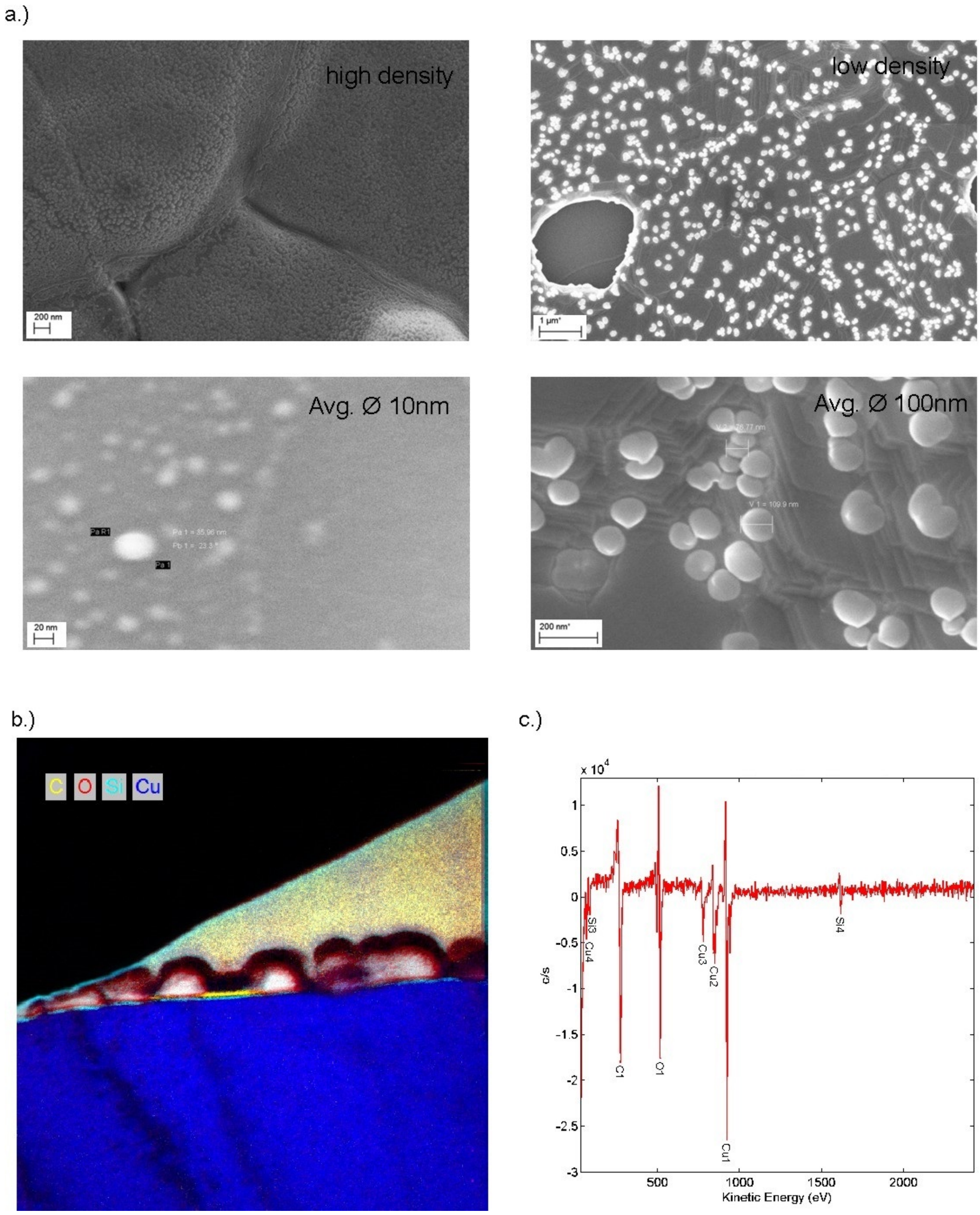

Figure 2: a) Growth conditions for the silicon nano-spheres lead to different sizes and coverage rates: Time and temperature are the most important parameters for the growth: Clusters from 10 - 100nm and coverage rates from 1 - 1000 clusters/$\mu m^2$ were observed. b) EFTEM (energy filtered transmission electron microscopy) Element mapping on a TEM cross-section after graphene growth on copper: There is no carbon/graphene apparent under the silicon nanospheres. The Si nanospheres are oxidized on the surface. Between the nanospheres a graphene sheet can be seen (marked in yellow). c) Auger electron spectrum of the copper/graphene surface coated with silicon nano-sphere.

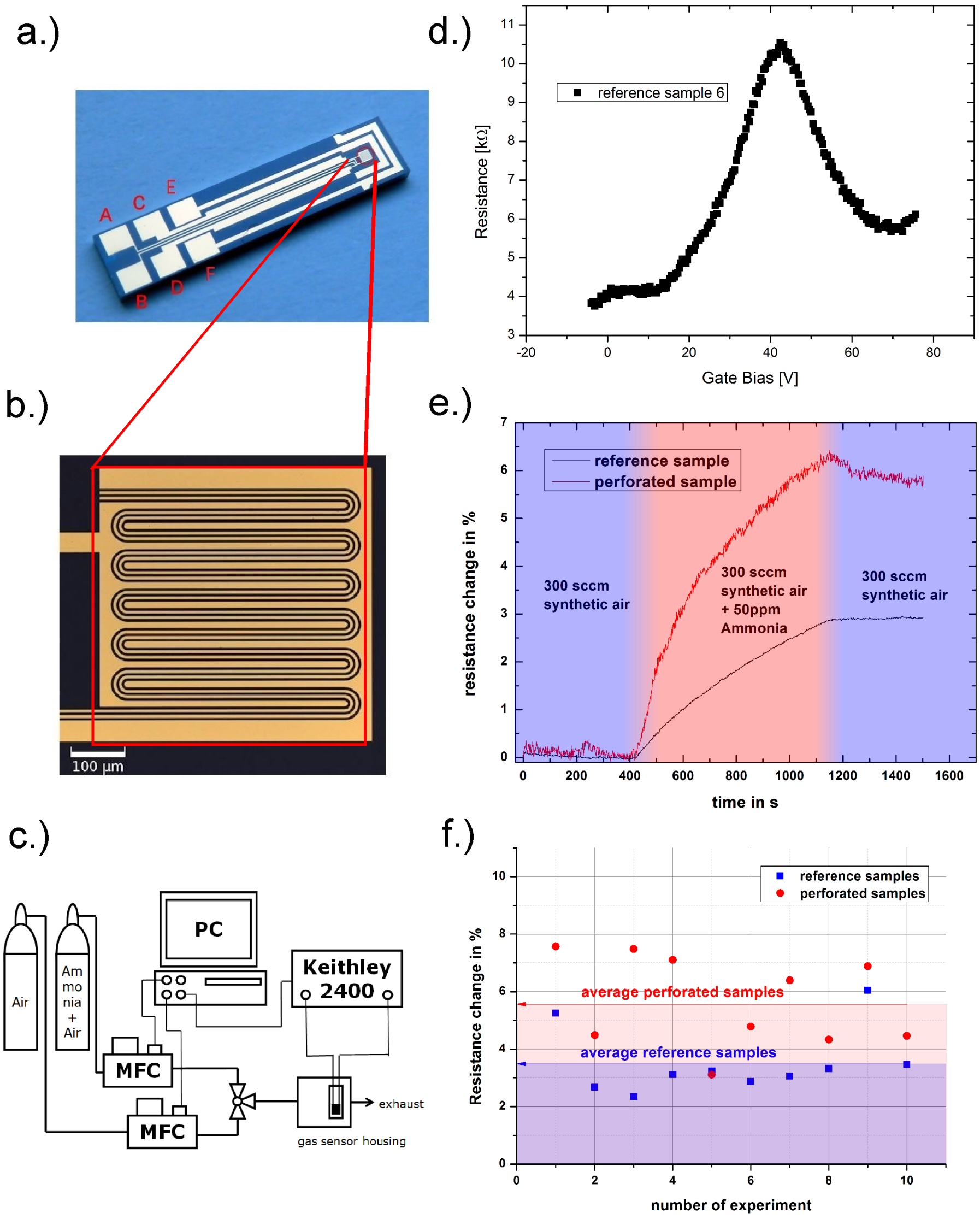

Figure 3 a) Optical micrograph of the test chip used for gas exposure experiments. Electrical measurements are carried out in a two-probe configuration using contacts A and B. The graphene reference and the graphene nanomesh were transferred onto the gold meander structure at the tip (indicated by red box). b) Gold meander structure and c) Schematic of measurement system. d) Back-Gate Sweep with device shown in a) (Dirac Point is visible). The voltage increase at voltages >70V can be attributed to a linear leakage current due to the very large device area. e) Electrical measurements of different samples exposed to a mixture of 50 ppm ammonia in synthetic air. The reference samples (ref) without perforation showed smaller resistance change than the perforated samples. f) Comparison of the resistance change of reference samples (average of 3.5%) and perforated samples (average of 5.6%).